\documentclass[final,5p,times,twocolumn]{elsarticle}

\usepackage{amssymb}

\usepackage{lineno}

\usepackage{upgreek}
\usepackage{subfigure}

\usepackage{wasysym} 

\newcommand{\comment}[1]{}
\newcommand{\ave}[1]{\langle#1\rangle}

\journal{Nucl. Instr. Meth. A}

\begin{document}

\begin{frontmatter}



\title{Status and Recent Results of the Acoustic Neutrino Detection Test System AMADEUS}


\author{R. Lahmann\corref{cor:1}}
\ead{robert.lahmann@physik.uni-erlangen.de}
\author{on behalf of the ANTARES Collaboration}

\cortext[cor:1]{Corresponding author; Tel.: +49\,9131\,8527147}

\address{Friedrich-Alexander-Universit\"{a}t
    Erlangen-N\"{u}rnberg, Erlangen Centre for Astroparticle Physics,
    Erwin-Rommel-Str. 1, 91058 Erlangen, Germany}

\begin{abstract}
The AMADEUS system
 is an integral part of the ANTARES neutrino
telescope in the Mediterranean Sea. The project aims at the
investigation of techniques for acoustic neutrino detection in the
deep sea. Installed at a depth of more than 2000\,m, the acoustic
sensors of AMADEUS are based on piezo-ceramics elements for the
broad-band recording of signals with frequencies ranging up to 125kHz.
AMADEUS was completed in May 2008 and comprises six ``acoustic
clusters'', each one holding six acoustic sensors that are arranged at
distances of roughly 1m from each other. The clusters are installed
with inter-spacings ranging from 15\,m to 340\,m.
Acoustic data are continuously acquired and processed at a computer
cluster where online filter algorithms are applied to select a
high-purity sample of neutrino-like signals. 1.6 TB of data were
recorded in 2008 and 3.2 TB in 2009. In order to assess the background
of neutrino-like signals in the deep sea, the characteristics of 
ambient noise and transient signals have been investigated.
In this article, the AMADEUS system will be described and recent
results will be presented.

\end{abstract}


\end{frontmatter}


\section{Introduction}
\label{sec:intro}

The AMADEUS project~\cite{bib:amadeus-2010} 
was conceived to perform a feasibility study for a
potential future large-scale acoustic neutrino detector. For this purpose, 
a dedicated array of acoustic sensors was integrated into the
ANTARES neutrino telescope~\cite{bib:ANTARES-paper, bib:ANTARES-line1}. 

Measuring acoustic pressure pulses in huge underwater acoustic arrays
is a promising approach for the detection of cosmic neutrinos with
energies exceeding 100\,PeV.  The pressure signals are produced by the
particle cascades that evolve when neutrinos interact with nuclei in
water.
The resulting energy deposition in a cylindrical volume of a few
centimetres in radius and several metres in length leads to a local
heating of the medium which is instantaneous with respect to the
hydrodynamic time scales.  This temperature change induces an
expansion or contraction of the medium depending on its volume
expansion coefficient.  According to the thermo-acoustic
model~\cite{bib:Askariyan2,bib:Learned}, the accelerated motion of the
heated volume---a micro-explosion---forms a pressure pulse of bipolar
shape which propagates in the surrounding medium.
Coherent superposition of the elementary sound waves, produced over the
volume of the energy deposition, leads to a propagation within a flat
disk-like volume (often referred to as {\em pancake})
in the direction perpendicular to the axis of the particle cascade.
After propagating several hundreds of metres in sea water, the pulse
has a characteristic frequency spectrum that is expected to peak
around 10\,kHz~\cite{bib:Sim_Acorne,bib:Sim_Acorne2,bib:Bertin_Niess}.
%
%

Two major advantages over an optical neutrino telescope motivate
studying acoustic detection.  First, the attenuation length in sea water
is about 5\,km (1\,km) for 10\,kHz (20\,kHz) signals.  This
is one to two orders of magnitude larger than for visible light
with a maximum attenuation length of about 60\,m.
The second advantage is the more compact sensor design and simpler readout
electronics for acoustic measurements. 
Since on the other hand the speed of sound
is small compared to the speed
of light, coincidence windows between two spatially separated sensors
are correspondingly large. Furthermore, the signal amplitude is
relatively small compared to the acoustic background in the sea,
resulting in a high trigger rate at the level of individual sensors
and making the implementation of efficient online data reduction
techniques essential.  To reduce the required processing time without
sacrificing the advantages given by the large attenuation length, the
concept of spatially separated clusters of acoustic sensors is used in
the AMADEUS system.  Online data filtering is then predominantly applied to the
closely arranged sensors within a cluster.

%

It is important to
realise that there are two kinds of background which need to be
understood
to assess the feasibility of an acoustic neutrino detector: 
First, there is ambient noise which can be described by
its characteristic power spectral density.  This noise is determined
by environmental processes and 
defines the minimum
pulse heights that can be measured, if a given signal-to-noise ratio
can be achieved with a search algorithm. 
Second, there are neutrino-like events, i.e. signals which have the
characteristic bipolar pulse shape 
but have a different origin. 
It is important to measure the spatial, temporal and pulse-height
distribution of such bipolar events
in order to asses the probability for random coincidences that
mimic the characteristic pancake structure of a neutrino sound wave.
For this kind of measurement, a hydrophone array is required and the
synchronisation among the hydrophones is crucial.


This article is organised as follows: In Sec.~\ref{sec:antares_detector}
an overview of the ANTARES neutrino telescope is given and in 
Sec.~\ref{sec:amadeus} the AMADEUS system is described.
In Sec.~\ref{sec:acoustic_prop} the 
properties at the ANTARES site are discussed which are responsible
for the sound propagation in the sea. In Sec.~\ref{sec:ambient-noise}  
first results of the measurement of the ambient noise are presented while in
Secs.~\ref{sec:pos_cali} and \ref{sec:source_dir_reco} the position 
calibration of the AMADEUS sensors and the direction reconstruction of 
acoustic sources, respectively, 
are discussed. These measurements are prerequisites to determine a density
of bipolar events in the Mediterranean Sea. In the subsequent sections, further
steps are discussed before conclusions are drawn.

\section{The ANTARES Detector}
\label{sec:antares_detector}

\begin{figure}[ht]
\centering
\includegraphics[width=8.0cm]{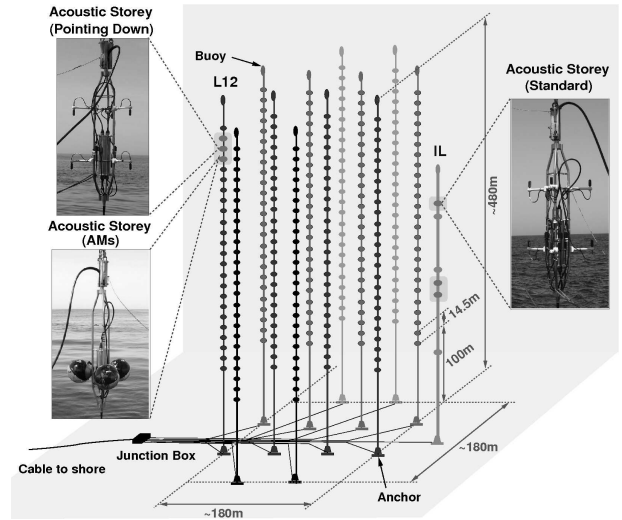}
\caption{A sketch of the ANTARES detector. 
The six acoustic storeys are highlighted and their three different setups
are shown (see text for details). 
L12 and IL denote the 12th detection line and the Instrumentation Line, 
respectively.}
\label{fig:ANTARES_schematic_all_storeys}
\end{figure}

The AMADEUS system~\cite{bib:amadeus-2010}
 is integrated into the ANTARES neutrino
telescope~\cite{bib:ANTARES-paper, bib:ANTARES-line1}, which was
designed to detect neutrinos by measuring the Cherenkov light emitted
along the tracks of relativistic secondary muons generated in neutrino
interactions.
A sketch of the detector, with the AMADEUS modules highlighted, is
shown in Figure~\ref{fig:ANTARES_schematic_all_storeys}.  The detector
is located in the Mediterranean Sea at a water depth of about 2500\,m,
roughly 40\,km south of the town of Toulon on the French coast 
at the geographic position of 42$^\circ$48$'$\,N, 6$^\circ$10$'$\,E.
It was
completed in May 2008 and
 comprises 12 vertical structures, the {\em detection lines}. 
Each detection line
holds up to 25 {\em storeys} that are arranged at equal distances of 14.5\,m
along the line, starting at about 100\,m above the sea
bed and interlinked by electro-optical cables.  A standard
storey consists of a titanium support structure, holding three {\em
  Optical Modules}~\cite{bib:OMs} 
(each one consisting of a photomultiplier tube (PMT) inside
a water-tight pressure-resistant glass sphere) and one {\em Local
  Control Module (LCM).}  
The LCM consists of 
a cylindrical titanium container
and the off-shore electronics within that container.
%
It comprises a {\em compass board}
that measures the tilt and the orientation
of the storey.
Timing correlations between storeys 
on a sub-nanosecond level are provided by a clock system that 
provides a master clock signal from a shore station, distributed
optically
to the individual LCMs over a dedicated set of fibres.

A 13th line, called {\em Instrumentation Line (IL)}, is equipped with
instruments for monitoring the environment. It holds six storeys.
For two pairs of consecutive storeys in the IL, the vertical distance
is increased to 80\,m.
Each line is fixed on the sea floor by an anchor equipped with
electronics and held taut by an immersed buoy.  An interlink cable
connects each line to the {\em Junction Box} from where the main
electro-optical cable provides the connection to the shore station.

The ANTARES lines are free to swing and twist in the undersea current.
In order to determine the positions of the storey with a precision of
about 20\,cm---which is necessary to achieve the required pointing
precision for neutrino astronomy---the detector is equipped with an
acoustic positioning system~\cite{bib:antares-pos}.  The system
employs an acoustic transceiver at the anchor of each line and four
autonomous transponders positioned around the 13 lines. Along each
detection line, five positioning hydrophones
receive the signals emitted by the transceivers.  By performing
multiple time delay measurements and using these to triangulate the
positions of the individual hydrophones,
the line shapes can be reconstructed relative to the positions of the
emitters. Currently, the sequence of positioning emissions is repeated
every 2 minutes.

\section{The AMADEUS System}
\label{sec:amadeus}
\subsection{Overview}
\begin{figure}[ht]
\centering
\hspace*{-10mm}\includegraphics[width=7.0cm]{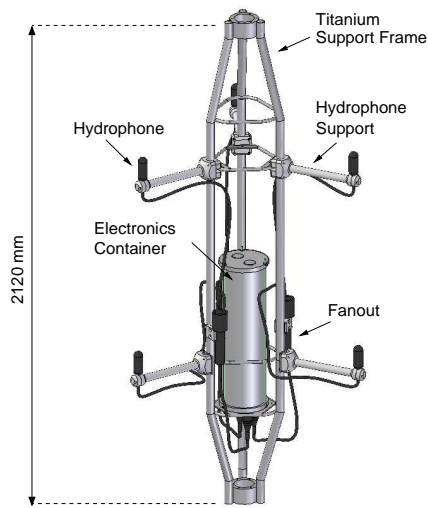}
\caption{
Drawing of a standard acoustic storey,  or acoustic cluster, 
with hydrophones.
\label{fig:acou_storey_drawing}
}
\end{figure}

In AMADEUS, acoustic sensing is integrated in the form of {\em
  acoustic storeys} that are modified versions of standard ANTARES
storeys, in which the Optical Modules are replaced by custom-designed
acoustic sensors.  
The acoustic storeys are
the implementation of the acoustic clusters discussed above.
The AMADEUS system uses the standard ANTARES facilities 
whenever possible, e.g.
the hardware and software for the 
data transmission to shore, the clock system, and the software
to operate the detector.
Dedicated electronics is used for the amplification, digitisation
and pre-processing of the analogue signals.  
Figure~\ref{fig:acou_storey_drawing} shows the design of a standard
acoustic storey with hydrophones.
Six acoustic sensors per storey were implemented, arranged at
distances of roughly 1\,m from each other. 
This number was the
maximum compatible with the design of the LCM and the bandwidth of
data transmission to shore.  
The data are digitised with 16 bit resolution and 250\,k samples per
second.

The AMADEUS system comprises a total of six acoustic storeys: three on
the IL, which started data taking in December 2007, and three on the
12th detection line (Line 12), which was connected to shore in May
2008.  AMADEUS is now fully functional and routinely taking data with
34 sensors. Two out of 36 hydrophones became inoperational during
their deployment.  

The acoustic storeys on the IL are located at 180\,m, 195\,m, and
305\,m above the sea floor.  On Line 12, which is anchored at a
horizontal distance of about 240\,m from the IL, the acoustic storeys
are positioned at heights of 380\,m, 395\,m, and 410\,m above the sea
floor.  
With this setup,
acoustic sensors are installed at water
depths between 2050 and 2300\,m
and the 
maximum distance between two acoustic
storeys is 340\,m.
AMADEUS hence covers three length scales: spacings of the order of
1\,m between sensors within a storey (i.e. an 
acoustic cluster); intermediate
distances of 14.5\,m between adjacent acoustic storeys within a
line; and large scales from about 100\,m vertical distance on the IL
up to 340\,m between storeys on different lines.
%
%
The AMADEUS system includes
time synchronisation and a continuously operating data acquisition
setup and is in principle scalable to a large-volume detector.


\subsection{Acoustic Sensors}
Two types of sensing devices are used in AMADEUS: hydrophones and {\em
  Acoustic Modules}~\cite{bib:amadeus-2010,bib:naumann_phd}. 
The acoustic sensors employ in both cases piezo-electric elements for
the broad-band recording of signals with frequencies ranging up to
125\,kHz. 
%
For the hydrophones, the piezo elements are coated in polyurethane,
whereas for the Acoustic Modules they are glued to the inside of
standard glass spheres which are normally used for Optical Modules.

The measurements presented here were done with the hydrophones;
The acoustic modules are described 
elsewhere~\cite{bib:amadeus-2010,bib:naumann_phd}. 
The typical sensitivity of the hydrophones is around
$-$145\,dB\,re\,1V/$\upmu$Pa (including preamplifier).
Different types of hydrophones are installed in AMADEUS, all of which
have a diameter of 38\,mm and
a length (from the cable junction to the opposite end) of 102\,mm. 
The equivalent inherent noise level in the frequency range from 1 to
50\,kHz is 
about 5.4\,mPa
for the AMADEUS hydrophones with the smallest such noise. 
This compares to 6.2\,mPa of the lowest
expected ambient noise level in the same frequency band for a
completely calm sea~\cite{bib:Graf_PhD_2008},
referred to as {\em sea state 0}~\cite{urick2}. 

\subsection{On-Shore Data Processing}
\label{sec:data_processing}
An on-shore computer cluster 
is used to process and filter the data stream and store the selected events.
%
The system is operating continuously and automatically, requiring only
little human intervention.
%
It currently consists of four
server-class computers, of which two with a total of 12 cores 
are used for data
triggering\footnote{While this functionality might be more commonly
  referred to as filtering, it is ANTARES convention to refer to the
  ``on-shore trigger''.}.  
%
%
One of the remaining two computers is used to write the data to an
internal 550 GB disk array, 
while the other is used to operate the
software for the online control of
the data acquisition and other miscellaneous 
processes and to provide
remote access to the system via the Internet.

The AMADEUS trigger searches the data by an adjustable software filter;
the events thus selected are stored to disk. This way the raw data
rate of about 1.5\,TB/day is reduced to about 10\,GB/day for storage.
Currently, three trigger schemes are in 
operation~\cite{bib:amadeus-2010,Neff_diplom}:
A minimum bias trigger which records data continuously for about 10\,s
every 60\,min, a threshold trigger which is activated when the signal
exceeds a predefined amplitude, and a pulse shape recognition
trigger. For the latter, a cross-correlation of the signal with a
predefined bipolar signal, as expected for a neutrino-induced cascade,
is performed. The trigger condition is met if the output of the
cross-correlation operation exceeds a predefined threshold.  With
respect to a matched filter, this implementation reduces the run time
complexity while yielding a comparable trigger performance.

In total, 1.6 TB of data were
recorded in 2008 and 3.2 TB in 2009.

\section{Properties of the Mediterranean Sea}
\label{sec:acoustic_prop}
\begin{figure}[ht]
\centering
\includegraphics[width=8.0cm]{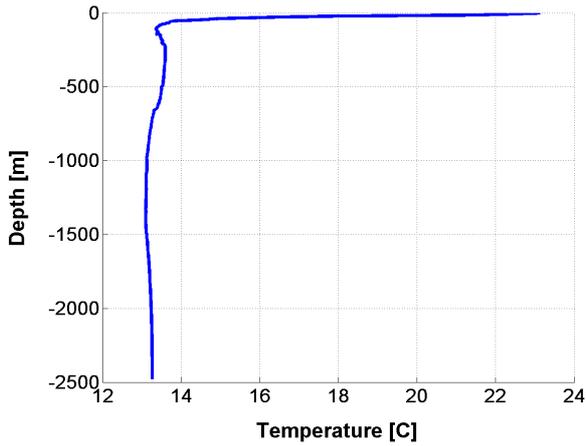}
\caption{Temperature as a function of the depth below the sea surface,
measured in a sea campaign near the ANTARES site in summer of 2007.
}
\label{fig:t_vs_d}
\end{figure}

The speed of sound in water depends on its temperature, salinity, and pressure 
(i.e. depth). 
%
A measurement of the temperature vs.\ the depth at the ANTARES site 
conducted in summer of 2007
is shown
in Fig.~\ref{fig:t_vs_d}. Only the upper $\sim$100\,m of the water
are affected by seasonal variations and 
below 
that depth, the temperature is quite stable, ranging from
13.1\,$^\circ$C to 13.6\,$^\circ$C.
Consequently, sound channelling is not a significant effect for acoustic
measurements at the ANTARES site. 
This is quite different from the situation in oceans, where typically 
the uppermost kilometre of water shows a temperature decrease with depth.

\begin{figure}[ht]
\centering
\includegraphics[width=8.0cm]{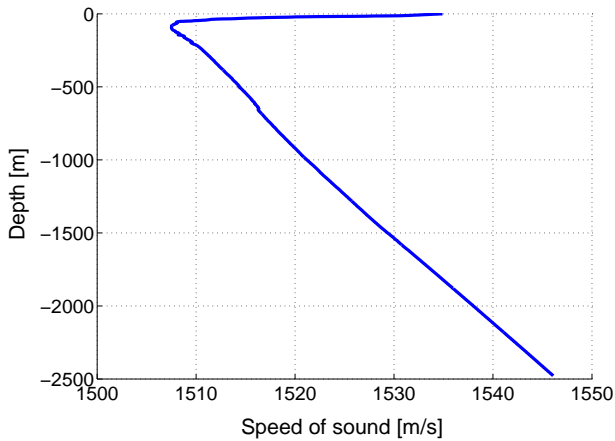}
\caption{Speed of sound as a function of the depth below the sea surface,
derived from the values of temperature, salinity and pressure measured
during a sea campaign near the ANTARES site in summer of 2007.
}
\label{fig:v_vs_d}
\end{figure}

The sound speed profile as a function of depth is shown in
Fig.~\ref{fig:v_vs_d}. Once the temperature is stable, the speed of
sound increases almost linearly with depth.  Assuming an open water
model, i.e. assuming that the temperature gradient and the depth of
the sea surrounding ANTARES do not vary, the furthest distance from
which an acoustic signal originating from the surface can reach the
AMADEUS device is about 30\,km. The refracted signal will reach the
uppermost sensor from an angle of about $-5.5^\circ$, i.e. from below a
horizontal plane. A signal originating at the surface and reaching the
detector within a horizontal plane will have a distance of about
20\,km. The open water model is an approximation that is not valid
for directions towards the coast. In northwards direction, for instance,
the water depth is reduced to about 200\,m  within roughly 10\,km.

\section{Ambient Noise}
\label{sec:ambient-noise}

To measure the ambient background present at the ANTARES site,
one sensor on the IL07 was evaluated from December 5, 2007 to January 22,
2010, i.e. over a period of about 2 years.  During this time, a total
of 18462 minimum bias samples (each one containing data continuously
recorded over a timespan of $\sim$10\,s) was recorded.

\begin{figure}[ht]
\centering
\includegraphics[width=8.0cm]{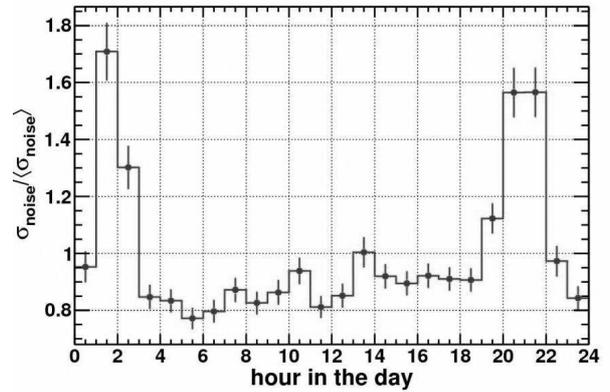}
\caption{
Ambient noise, measured as described in the text, as a function of the time
of the day. The noise is shown relative to the mean noise recorded over the 
complete period of about two years that was used for the analysis.
}
\label{fig:ambient_noise_over_day}
\end{figure}

Removing samples with large components at high frequencies
(e.g. containing signals from the acoustic positioning system and non-Gaussian 
distributions) 13909 samples
(75.4\%) are remaining.  For each of these samples, the noise power
spectral density (PSD) was integrated in the frequency range $f = 1 - 50$\,kHz.
The resulting noise values, relative to the mean noise
over all samples, are shown in Fig.~\ref{fig:ambient_noise_over_day}
as a function of the time of the day. Two clear peaks can be observed
at about 2 a.m. and 9 p.m.; while the origin of this structure can not
yet be undoubtedly determined, the most likely source is shipping traffic.

\begin{figure}[ht]
\centering
\includegraphics[width=8.0cm]{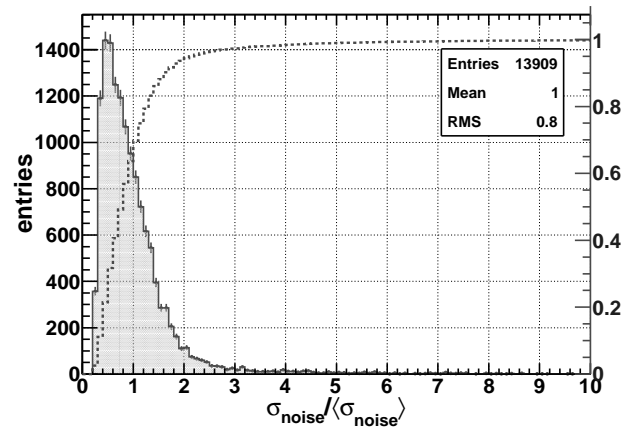}
\caption{
Frequency of occurrence distribution
for the ambient noise, relative to the mean ambient noise 
  recorded over the 
complete period of about two years that was used for the analysis
(left scale, filled histogram).
Also shown is the accumulative distribution, normalised to
the total number of entries of the distribution (right scale, solid 
line). 
}
\label{fig:noise_distr}
\end{figure}

Figure~\ref{fig:noise_distr} shows the frequency of occurrence distribution of 
the same values. Also shown is the corresponding accumulative distribution. 
For
95\% of the samples, the noise level is below $2\ave{\sigma_{noise}}$,
demonstrating that the ambient noise conditions are very stable.

All sensors have been calibrated in the laboratory prior to
deployment.
The absolute noise level can be estimated by assuming
a constant sensor sensitivity\footnote{
The ambient noise is originating mainly from the sea surface and hence
displays a directivity which has to be folded with the
variations of the sensitivity over the polar angle to obtain an effective
average sensitivity. 
For the results presented here, the noise has been assumed to be isotropic.
}
of $-145\pm2$\,dB\,re\,1V/$\upmu$Pa. 
With this value, the mean noise level is 
$\ave{\sigma_{noise}} = 25^{+7}_{-5}$\,mPa with the median of the distribution
at $0.7 \ave{\sigma_{noise}} \sim 17$\,mPa. 
It must be understood that these noise levels correspond to the specific 
frequency range chosen. The frequency range used for this study is
{\em not} optimised for the spectral shape of the expected neutrino signal.
Clearly, the noise level can be made arbitrarily small by
choosing a correspondingly small frequency range. The most sensible
procedure is to choose the frequency range that optimises the signal-to-noise 
ratio (SNR) for the expected neutrino signal. Preliminary studies
using the shower 
parametrisation and algorithms presented in~\cite{bib:Sim_Acorne2} 
indicate that this range is about 10 to 50\,kHz. 
The AMADEUS minimum bias data show that the 
RMS noise of 25\,mPa for $f = 1-50$\,kHz reduces to about 7.5\,mPa 
for $f = 10-50$\,kHz, i.e. by a factor of about 0.3. 
The estimate for the 
energy sensitivity of the AMADEUS device given below,
based on the frequency range of  $1-50$\,kHz,
therefore has to be understood as a
conservative upper bound. 
A more realistic energy sensitivity is obtained by applying 
the factor
of 0.3 to scale the results to the smaller frequency range.

Currently, the detection threshold for bipolar signals corresponds to
a SNR of about two for an individual hydrophone.
By applying pattern recognition methods that are more closely tuned to the
expected neutrino signal, this threshold can possibly be reduced.
For a SNR of two, the noise value of $2\ave{\sigma_{noise}}$ below which 
$95$\% of the data can be found corresponds to a signal amplitude 
of 100\,mPa, which corresponds to a neutrino energy of $\sim$10 EeV at a 
distance of 200\,m~\cite{bib:Sim_Acorne}. 
Scaling this result to the reduced frequency range,
and using the median noise level of $0.7\ave{\sigma_{noise}}$ it can be 
estimated that for 50\% of the time, the energy threshold is at $\sim$1 EeV
or below 
for a distance of 200\,m. 

In summary, the ambient noise conditions are very favourable for acoustic
neutrino detection in the Mediterranean Sea: The noise level is stable 
at an expected level.
Hence it will be crucial to determine the density of 
bipolar events to assess the feasibility of an acoustic neutrino detector.
A prerequisite for the position reconstruction of acoustics sources is the
position calibration of the acoustic storeys which will now be described.

\section{Position Calibration}
\label{sec:pos_cali}
Just as for the PMTs in the standard storeys, the relative positions
of the acoustic sensors within the detector have to be continuously
monitored.  This is done by using the emitter signals of the ANTARES
acoustic positioning system.
%
The emission times of the
positioning signals are recorded in the central ANTARES database.
Measuring the reception time within the acoustic sensors of 
the AMADEUS system allows for the calculation of the distance between
emitter and receiver.
%
Using the signals from multiple
emitters and the knowledge of their positions at the anchors of the lines, the
positions of the AMADEUS sensors can be reconstructed.
And finally, 
from the reconstructed positions of the individual sensors, the six
degrees of freedom (centre of mass coordinates and the three Euler angles) 
of each storey can be obtained.

\begin{figure}[ht]
\centering
\includegraphics[width=9.0cm]{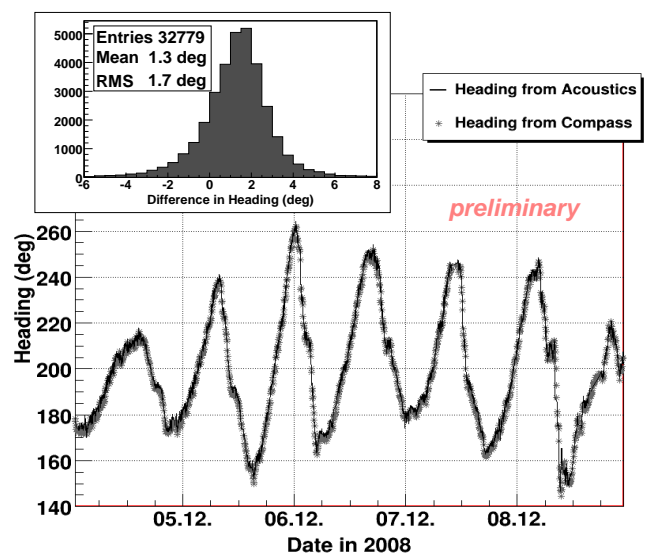}
\caption{
Comparison of the heading measured with the sensors of the uppermost
storey of IL07 and with the compass board in the same storey over a period 
of five days. In the insert at the top left, the resolution between
the two measurements is shown. 
}
\label{fig:compass_vs_acoustics}
\end{figure}

Figure~\ref{fig:compass_vs_acoustics} shows the comparison of the 
{\em heading}
as measured with the acoustic sensors on a storey and the compass board
installed in that storey. The heading 
is defined as 
the rotational angle of the storey around the vertical axis, relative
to a well defined reference horizontal axis within the storey. 
A heading of zero 
degrees then corresponds to the reference axis pointing north,
with angles counting clockwise from $0^\circ$ to $360^\circ$.
The periodic structure of the heading is due to the sea currents which change
in accordance to the Coriolis force, which at a geographic latitude $\phi$
affects the direction of sea currents with a period of 
$T = T_{0}\,\sin\phi$,
where  $T_{0}\approx 24$\,h is the period of a sidereal day on Earth 
and therefore $T \approx 16.3$\,h.

Both the acoustic measurement and the measurement of the compass board
have an offset of their zero degree position w.r.t. the cardinal direction
of north: When measuring zero degrees, the compass board points to
the North Magnetic Pole while the direction derived from the acoustic 
measurements points along the {\em northing axis} of the UTM grid. At the
location of ANTARES, this difference is 2.5$^\circ$ where the UTM
northing axis points further westwards than the direction of the North
Magnetic Pole. The difference shown in Fig~\ref{fig:compass_vs_acoustics}
is not corrected for these offsets, the mean value of the distribution after
this correction is therefore $-1.2^\circ$. The offsets from the six storeys 
have a spread of about $1^\circ$ around zero.

The RMS deviation between the two measurements is 1.7$^\circ$ (0.03\,rad)
or 3\% w.r.t. the distance of a source from the line. Here the
residual offset between compass board and angle reconstruction with the
AMADEUS sensors were ignored. 
On the other hand, the RMS deviation is the quadratic sum of the
resolution of the compass board and of the AMADEUS measurement,
resulting in an overestimation of the RMS resolution. 
Since the two effects are
of about the same size, they will roughly cancel out.

For a measurement of the density of bipolar events, the absolute
positioning of the acoustic storeys is not important. The precision of
the measurement of source positions (which then determines the error of
the density and limits the distance up to which a surveillance is
possible) is determined by the resolution of the angular reconstruction
of a source position and by that of the distance between the
storeys. The horizontal distance between the IL07 and Line 12 is about
240m. Preliminary studies, comparing the positions that are
reconstructed with AMADEUS and with the ANTARES positioning system,
indicate that the distance between the centre of mass  positions of two storeys
on the two lines can be reconstructed with a precision of much better
than 1\,m. The resulting precision of $\Delta(\,\vec{r}_{L12} -
\vec{r}_{IL07}\,)/ |\,\vec{r}_{L12} - \vec{r}_{IL07}\,| \ll$ 1\%
indicates that the dominant contribution to the error on the position
reconstruction will stem from the angular resolution of the source
position reconstruction (see the following section).

\section{Source Direction Reconstruction}
\label{sec:source_dir_reco}
The sensors within a cluster allow for efficient triggering of
transient signals and for direction reconstruction.  The combination
of the direction information from different acoustic storeys yields
(after verifying the consistency of the signal arrival times at the
respective storeys) the position of an acoustic
source~\cite{bib:Richardt_reco}.
\begin{figure}[ht]
\centering
\includegraphics[width=9.0cm]{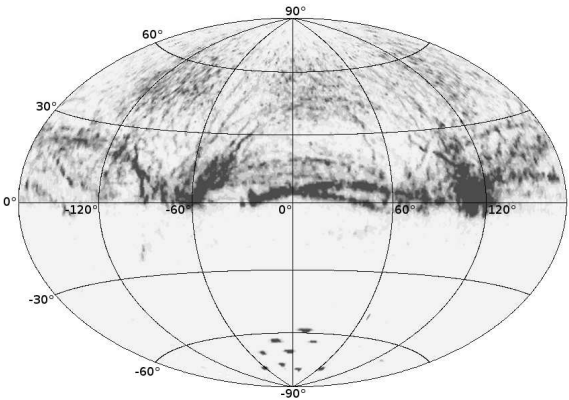}
\caption{
Map of directions of sources as reconstructed with an acoustic storey on
Line 12. Zero degrees in azimuth correspond to the north direction, the 
polar angle of zero corresponds to the horizon of an observer on the
acoustic storey.
}
\label{fig:skymap}
\end{figure}
Figure~\ref{fig:skymap} shows the reconstructed directions of all
sources that were triggered during a period of one month. 
One can observe dark bands of increased acoustic activity. These bands 
can be associated with shipping routes and the points with 
high activities with the directions of ports. Note that 
in accordance with the arguments given in Sec.~\ref{sec:acoustic_prop}, the
ports themselves are too far away to record acoustic signals directly, 
unless reflection at 
the sea bottom and surface occur, which would then however strongly 
reduce the  signal strength.
It is obvious from Fig.~\ref{fig:skymap} that a fiducial volume for the 
determination of the background rate of bipolar events must exclude the 
sea surface.

The resolution of the reconstruction is about 3$^\circ$ in azimuth ($\phi$)
and 0.5$^\circ$ in the polar angle ($\theta$)~\cite{bib:richardt_phd}. 
While the source 
direction reconstruction uses methods similar to those used for the 
position reconstruction with AMADEUS, for the latter typical $7\sim8$ emitters
are used which increases the statistics and hence the precision. Note
that for both methods the error includes the resolution of the compass board.

\section{Further Steps}


Using the direction reconstruction from several storeys, source position
reconstruction  can be performed. First investigations over a 
three month period show that a detailed classification of events
triggered by the  
pulse shape recognition trigger (see Sec.~\ref{sec:data_processing})
is necessary to reduce the background
of neutrino-like events~\cite{bib:richardt_phd}. 
Several methods for this signal classification are
currently under investigation~\cite{bib:Neff_BIPselection}.

Furthermore, Monte Carlo code is under development, based on the software
and method presented in~\cite{bib:Sim_Acorne2}. This will allow for testing
the classification and calculating efficiencies for the classification of 
neutrino events.

\section{Conclusions}
Recent results from the acoustic neutrino detection test system AMADEUS have
been presented. The measurement of the ambient noise over a period
of about two years shows that the noise level is very stable and at the 
expected level, allowing for measurements of neutrino energies down to 
$\sim$1\,EeV.
%
The AMADEUS system has all features of an acoustic neutrino telescope,
except for its size. As would be necessary for a potential future
acoustic neutrino telescope in water, position calibration of its
acoustic sensors, which are swaying with the underwater sea currents,
are routinely performed with AMADEUS.
For individual acoustic storeys,
comprising six acoustic sensors each,    
reconstruction of acoustic source directions is done with an 
angular resolution of about $3^\circ$ in azimuth.

Current focus of the analysis work is the further classification of transient
bipolar events to minimise the irreducible background for neutrino searches. 
In addition, work on MC simulations is in progress to test these classification
methods and to calculate efficiencies for neutrino detection.

In summary,
the system is excellently suited 
to assess the background conditions for the measurement of the
bipolar pulses expected to originate from neutrino interactions. 
The AMADEUS system can furthermore be used as a multi purpose device to perform
positioning and marine science investigations in addition to 
studies of neutrino detection.

\section{Acknowledgements}
This study was supported by the German government through BMBF grants
5CN5WE1/7 and 05A08WE1. Furthermore, the author wishes to thank the organisers 
of the ARENA 2010 workshop for an enjoyable and instructive event.





\bibliographystyle{elsarticle-num}
\bibliography{arena2010_lahmann}

\begin{thebibliography}{10}
\expandafter\ifx\csname url\endcsname\relax
  \def\url#1{\texttt{#1}}\fi
\expandafter\ifx\csname urlprefix\endcsname\relax\def\urlprefix{URL }\fi
\expandafter\ifx\csname href\endcsname\relax
  \def\href#1#2{#2} \def\path#1{#1}\fi

\bibitem{bib:amadeus-2010}
{J.A. Aguilar \etal\ (ANTARES Coll.)}, 
 {doi:10.1016/j.nima.2010.09.053},
{to be
  published by Nucl.\ Instr.\ and Meth. A}, arXiv:1009.4179v1 [astro-ph.IM].

\bibitem{bib:ANTARES-paper}
{The ANTARES Collaboration},
 ANTARES, the first operational Neutrino Telescope in the Mediterranean
Sea,
to be submitted to Nucl.\ Instr.\ and Meth.\ A.

\bibitem{bib:ANTARES-line1}
{M.\ Ageron et al.\ (ANTARES Coll.)}, 
Astropart.\ Phys. 31 (2009) 277, arXiv: 0812.2095 v1 [astro-ph].

\bibitem{bib:Askariyan2}
{G.A.\ Askariyan, B.A.\ Dolgoshein \etal}, 
Nucl.\ Instr.\ and Meth. 164 (1979) 267.

\bibitem{bib:Learned}
J.~Learned, 
Phys. Rev. D 19 (1979) 3293.

\bibitem{bib:Sim_Acorne}
{S.~Bevan \etal\ (ACoRNE Coll.)}, 
Astropart.\ Phys. 28~(3) (2007) 366,
  arXiv:astro-ph/0704.1025v1.

\bibitem{bib:Sim_Acorne2}
{S.~Bevan \etal\ (ACoRNE Coll.)}, 
Nucl.\ Instr.\ and Meth. A 607 (2009)
  389, arXiv:0903.0949v2 [astro-ph.IM].

\bibitem{bib:Bertin_Niess}
{V.\ Niess and V.\ Bertin}, 
Astropart.\ Phys. 26 (2006) 243, arXiv:astro-ph/0511617v3.

\bibitem{bib:OMs}
{P.~Amram \etal\ (ANTARES Coll.)}, 
Nucl.\ Instr.\ and Meth. A 484 (2002) 369.

\bibitem{bib:antares-pos}
{M.\ Ardid for the ANTARES Coll.}, 
in: Proceedings of VLVnT 2008, the 3rd International Workshop on a
  Very Large Volume Neutrino Telescope for the Mediterranean Sea, Vol. A 602,
  2009, p. 174.

\bibitem{bib:naumann_phd}
{C.L.~Naumann}, Development of sensors for the acoustic detection of ultra high
  energy neutrinos in the deep sea, Ph.D. thesis, Univ.\ Erlangen-N{\"u}rnberg,
  2007, FAU-PI4-DISS-07-002.

\bibitem{bib:Graf_PhD_2008}
{K.~Graf}, Experimental studies within ANTARES towards acoustic detection of
  ultra-high energy neutrinos in the deep sea, Ph.D. thesis, Univ.\
  Erlangen-N{\"u}rnberg, 2008, FAU-PI1-DISS-08-001.

\bibitem{urick2}
{R.J.\ Urick}, Ambient Noise in the Sea, Peninsula publishing, Los Altos, USA,
  1986, {ISBN 0-932146-13-9}.

\bibitem{Neff_diplom}
M.~Neff, {Studie zur akustischen Teilchendetektion im Rahmen des
  ANTARES-Experiments: Entwicklung und Integration von Datennahmesoftware},
  Diploma thesis, Univ.\ Erlangen-N{\"u}rnberg, 2007, FAU-PI1-DIPL-07-003.

\bibitem{bib:Richardt_reco}
C.~Richardt\ \etal, 
Astropart.\ Phys. 31 (2009) 19, arXiv:0906.1718v1 [astro-ph.IM].

\bibitem{bib:richardt_phd}
{C. Richardt}, {Acoustic particle detection - Direction and source location
  reconstruction techniques}, Ph.D. thesis, Univ.\ Erlangen-N{\"u}rnberg,
  2010, FAU-ECAP-DISS-10-003.

\bibitem{bib:Neff_BIPselection}
{M. Neff}, these proceedings.

\end{thebibliography}







\end{document}